\documentstyle[aps,floats,epsf,psfig,fleqn]{revtex}

\DeclareFontFamily{U}{msb}{}
\DeclareFontShape{U}{msb}{m}{n}{
<5><6><7><8><9> gen *msbm <10><10.95><12><14.4><17.28><20.74><24.88>msbm10}{}
\DeclareSymbolFont{AMSb}{U}{msb}{m}{n}
\DeclareMathSymbol{\AAA}{\mathbin}{AMSb}{'101}
\DeclareMathSymbol{\BBB}{\mathbin}{AMSb}{'102}
\DeclareMathSymbol{\CCC}{\mathbin}{AMSb}{'103}
\DeclareMathSymbol{\DDDD}{\mathbin}{AMSb}{'104}
\DeclareMathSymbol{\EEE}{\mathbin}{AMSb}{'105}
\DeclareMathSymbol{\FFF}{\mathbin}{AMSb}{'106}
\DeclareMathSymbol{\GGG}{\mathbin}{AMSb}{'107}
\DeclareMathSymbol{\HHH}{\mathbin}{AMSb}{'110}
\DeclareMathSymbol{\III}{\mathbin}{AMSb}{'111}
\DeclareMathSymbol{\JJJ}{\mathbin}{AMSb}{'112}
\DeclareMathSymbol{\KKK}{\mathbin}{AMSb}{'113}
\DeclareMathSymbol{\LLL}{\mathbin}{AMSb}{'114}
\DeclareMathSymbol{\MMM}{\mathbin}{AMSb}{'115}
\DeclareMathSymbol{\NNN}{\mathbin}{AMSb}{'116}
\DeclareMathSymbol{\OOO}{\mathbin}{AMSb}{'117}
\DeclareMathSymbol{\PPPP}{\mathbin}{AMSb}{'120}
\DeclareMathSymbol{\QQQ}{\mathbin}{AMSb}{'121}
\DeclareMathSymbol{\RRR}{\mathbin}{AMSb}{'122}
\DeclareMathSymbol{\SSS}{\mathbin}{AMSb}{'123}
\DeclareMathSymbol{\TTTTT}{\mathbin}{AMSb}{'124}
\DeclareMathSymbol{\UUU}{\mathbin}{AMSb}{'125}
\DeclareMathSymbol{\VVVV}{\mathbin}{AMSb}{'126}
\DeclareMathSymbol{\WWW}{\mathbin}{AMSb}{'127}
\DeclareMathSymbol{\XXXX}{\mathbin}{AMSb}{'130}
\DeclareMathSymbol{\YYY}{\mathbin}{AMSb}{'121}
\DeclareMathSymbol{\ZZZZ}{\mathbin}{AMSb}{'132}

\newcommand{\ds}{\displaystyle}

\newcommand{\wa}{{\scriptscriptstyle{\cal W}}}
\newcommand{\oi}{{\scriptscriptstyle{\cal O}}}

\newcommand{\dip}{{\alpha}}
\newcommand{\kwr}{k_\wa^r}
\newcommand{\kor}{k_\oi^r}

\newcommand{\Sor}{S_{\oi r}}
\newcommand{\Swi}{S_{\wa i}}

\newcommand{\MK}{{\bf K}}
\newcommand{\ML}{{\bf L}}
\newcommand{\MR}{{\bf O}}

\newcommand{\Ca}{{\rm Ca}}

\newcommand{\Bo}{{\rm Bo}}

\newcommand{\ggg}{\widehat{\bf g}}

\newcommand{\MAA}{\widehat{\bf A}}
\newcommand{\MKK}{\widehat{\bf K}}
\newcommand{\MLL}{\widehat{\bf L}}

\newcommand{\CA}{\overline{\rm Ca}}
\newcommand{\GR}{\overline{\rm Gr}}
\newcommand{\GL}{\overline{\rm Gl}}

\newcommand{\ph}{{\phi}}

\newcommand{\SS}{{S}}
\newcommand{\TT}{{t}}

\newcommand{\PP}{{P}}
\newcommand{\LL}{{l}}
\newcommand{\KK}{{k}}

\newcommand{\XX}{{\bf x}}
\newcommand{\EE}{{\bf e}}
\newcommand{\NAB}{{\mbox{\boldmath $\nabla$}}}

\newcommand{\TTT}{\widehat{{t}}}

\newcommand{\PPP}{\widehat{{P}}}
\newcommand{\XXX}{\widehat{{\bf x}}}
\newcommand{\NABB}{\widehat{{\mbox{\boldmath $\nabla$}}}}

\begin{document}
\draft
\setcounter{page}{0}
\title{
Trapping and Mobilization of Residual Fluid During Capillary
Desaturation in Porous Media}
\author{A. Lucia$\mbox{\rm n}^{1,2}$ and R. Hilfe$\mbox{\rm r}^{2,3}$}
\address{
$\mbox{ }^1$Inst. Atomic Physics, I.N.F.L.P.R., Lab. 22, P.O Box MG-7
76900 Bukarest, Romania\\
$\mbox{ }^2$ICA-1, Universit{\"a}t Stuttgart,
Pfaffenwaldring 27, 70569 Stuttgart\\
$\mbox{ }^3$Institut f{\"u}r Physik,
Universit{\"a}t Mainz,
55099 Mainz, Germany}
\maketitle
\thispagestyle{empty}
\begin{abstract}
We discuss the problem of trapping and mobilization of nonwetting
fluids during immiscible two phase displacement processes in porous
media.
Capillary desaturation curves give residual saturations as a function of
capillary number.
Interpreting capillary numbers as the ratio of viscous to capillary
forces the breakpoint in experimental curves contradicts the 
theoretically predicted force balance.
We show that replotting the data against a novel macroscopic capillary
number resolves the problem for discontinuous mode displacement.
\end{abstract}
\vspace{.7cm}
\newpage
\section{Introduction}
A displacement of one fluid by another within a porous
medium poses challenging problems of micro-to-macro-scale 
transitions which have received 
considerable attention from physicists in recent years
\cite{BK90b,oxa91,OBFJMA91,hil91d,hil92a,hil92b,BKS92,FFJM92,MFFJ92,KHT92,NHH92,sah93,BS95a,HS95,sch95,FMF96,hil96c,KKT97b,BS97,AAMHSS97}.
For reviews the reader is referred to \cite{sah93,hil95d}.
Apart from the omnipresent quenched correlated disorder
and structural heterogeneity, fluid-fluid and fluid-solid
interactions generate metastability and hysteresis
phenomena which have resisted quantitative 
prediction and understanding.
A central problem of great practical importance is the
prediction of residual nonwetting fluid saturation
after flooding the pore space with immiscible wetting
fluid \cite{wil86,hil98b,BH98}. Obvious interest in this problem
arises from enhanced oil recovery \cite{lak89,OBP92} 
or in situ remediation of soil contaminants \cite{STA93,hel97}. 

Microscopically the laws of hydrodynamics governing the 
pore scale processes are well known.
The complexity of the microscopic fluid movements and the 
lack of knowledge about the microstructure and wetting 
properties, however, renders a detailed microscopic 
treatment impossible.
Instead one has to resort to a more macroscopic treatment.

Desaturation experiments \cite{DB54,tab69,TKS73,abr75,CKM88,WM85,LDS81,MCT88}
show that the conventional macroscopic description is incomplete
\cite{hil98b,BH98}.
We shall begin our discussion by reminding the readers 
of this incompleteness.
We then exhibit another problem \cite{hil94c}.
This arises from the fact that microscopic capillary numbers 
seemingly cannot represent the force balance in a desaturation 
experiment.

Given these problems our main objective is to show that 
the recent analysis of \cite{hil94c} 
gives the correct force balance between 
macroscopic viscous and capillary forces for continuous
mode desaturation experiments.
For so called discontinuous mode displacements it leads 
to bounds on the size of residual blobs.

\section{Problems of two-phase flow equations}

Let us begin our discussion with the standard macroscopic
equations of motion for two phase immiscible displacement.
Macroscopic equations of motion describe multiphase flow on 
length scales large compared to a typical pore diameter.
Hence they are applied to laboratory samples with linear
dimensions on the order of centimeteres 
as well as to whole reservoirs measuring kilometers and more.
Consider a system having lengths
$L_x,L_y,L_z$ in the three spatial directions.
The pore space is assumed to be filled with two 
immiscible fluids denoted generically as water (index $\wa$)
and oil (index $\oi$).
The equations read \cite{sch74,dul92,sah95,hil94c}
\begin{equation}
\ph \frac{\ds \partial \SS_\wa}{\ds \partial \TT} =
\NAB \cdot \left\{
\frac{\ds \MK\;\kwr}{\ds \mu_\wa}
\left[\NAB \PP_\wa -\rho_\wa g\:\MR\:\EE\right]
\right\}
\label{Em5}
\end{equation}
\begin{equation}
-\ph \frac{\ds \partial \SS_\wa}{\ds \partial \TT} =
\NAB \cdot \left\{\frac{\ds \MK\;\kor}{\ds \mu_\oi}
\left[\NAB (\PP_\wa + \PP_c) -\rho_\oi g\:\MR\:\EE\right]
\right\}
\label{Em6}
\end{equation}
and they are supplemented with the constitutive relationships
\begin{eqnarray}
\kwr(\XX,\TT) & = & \kwr(\SS_n(\XX,\TT))\\
\kor(\XX,\TT) & = & \kor(\SS_n(\XX,\TT))\\
\PP_c(\XX,\TT) & = & \PP_c(\SS_n(\XX,\TT))
\end{eqnarray}
where
\begin{equation}
S_n(\XX,\TT) = \frac{\SS_\wa(\XX,\TT)-\Swi}{1-\Swi-\Sor} .
\label{Em2}
\end{equation}
The variables in these equations are the pressure field of
water denoted as $\PP_\wa$, and the water saturation $\SS_\wa$.
The saturation is defined as the ratio of water volume to pore space
volume.
Pressures and saturations are averages over a macroscopic region 
much larger than the pore size, but much smaller than the system size.
Their arguments are the macroscopic space and time variables $(\XX,\TT)$. 
The saturations obey $\Swi < \SS_\wa < 1-\Sor$ where the two
numbers $0\leq\Swi,\Sor\leq 1$ are two parameters representing the
irreducible water saturation, $\Swi$, and the residual oil
saturation, $\Sor$.
The residual oil saturation gives the amount of oil remaining
in a porous medium after water injection.
The normalized saturation $\SS_n$ varies between $0$ and $1$
as $\SS_\wa$ varies between $\Swi$ and $\Sor$.
The permeability of the porous medium is given by the
absolute (single phase flow) permeability tensor $\MK$.
The porosity $\ph$ is the volume fraction of pore space.
The two fluids are characterized by their viscosities $\mu_\wa$,
$\mu_\oi$ and their densities $\rho_\wa$,$\rho_\oi$.
The terms $\rho g\:\MR\:\EE$ represent the gravitational body 
force where $\EE^T=(0,0,-1)$ is a unit row vector pointing along
the negative $z$-axis, and $g$ is the acceleration of gravity.
The orthogonal matrix 
\begin{equation}
\MR =
\left(
  \begin{array}{ccc}
   \cos \dip_y & 0 & \sin \dip_y\\
   0 & 1 & 0\\
   -\sin \dip_y & 0 & \cos \dip_y
  \end{array}
\right)
\end{equation}
describes an inclination of the system. Here a rotation around the
$y$-axis with tilt or dip angle $\dip_y$ was assumed.
The  macroscopic capillary pressure $\PP_c$ is defined as the pressure 
difference between the oil and the water phase.
The constitutive relation for $\PP_c$ assumes that the  capillary
pressure function $\PP_c$ depends only on the saturation \cite{bea72}.
In addition to $\PP_c$ the dimensionless relative permeabilities are assumed
to be functions of saturation only, $\kwr=\kwr(S_\wa)$,$\kor=\kor(S_\wa)$.
They represent the reduction in permeability for one phase due to
the presence of the other phase \cite{sch74,dul92}.
The three constitutive relations
$\kwr(\SS_\wa),\kor(\SS_\wa)$ and $\PP_c(\SS_\wa)$
are assumed to be known from experiment.
Equations (\ref{Em5}) and (\ref{Em6})
are coupled nonlinear partial differential equations which
must be complemented with large scale boundary
conditions. For laboratory experiments the boundary
conditions are typically given by a surface source 
on one side of the sample, a surface sink on the
opposite face, and impermeable walls on the other faces.
For a geosystem the boundary conditions depend upon
the well configuration and the geological modeling of the
reservoir environment.

The problem with the macroscopic equations of motion
(\ref{Em5})-(\ref{Em6}) arises from the experimental
observation that the parameters $\Sor$ and $\Swi$
are not constant and known, but depend strongly
on the flow conditions in the experiment
\cite{DB54,tab69,TKS73,abr75,CKM88,WM85,LDS81,MCT88,OBP92}.
Hence they may vary in space and time.
More precisely, the residual oil saturation depends
strongly on the microscopic capillary number
$\Ca=\mu_\wa v/\sigma_{\oi\wa}$ where $v$ is a
typical flow velocity and $\sigma_{\oi\wa}$ is
the surface tension between the two fluids.
The capillary desaturation curve $\Sor(\Ca)$  is shown in 
Figure \ref{ocp} for unconsolidated glass beads and
sandstones \cite{abr75,MCT88}.
Such curves contradict clearly to the assumption that
the functions $\kwr(\SS_\wa),\kor(\SS_\wa)$ and $\PP_c(\SS_\wa)$
depend only on saturation.
Instead they show that $\kwr(\SS_\wa),\kor(\SS_\wa)$ and $\PP_c(\SS_\wa)$ 
depend also on velocity \cite{deg88,kal92} and pressure.
Hence they depend on the solution and cannot be considered to be 
constitutive relations characterizing the system.
The dependence shows that the system of equations of motion
is incomplete.

The breakpoint in a capillary desaturation curve marks
the point where the viscous forces, which attempt to mobilize
the oil, become stronger than the capillary forces, which
try to keep the oil in place.
For this reason the capillary desaturation curves are
usually plotted against $\Ca$ which represents the 
microscopic force balance between viscous and capillary
forces.
From Figure \ref{ocp} it is seen that the theoretical
force balance corresponding to $\Ca=1$ on the abscissa and
the experimental force balances represented by the various
breakpoints at $\Ca\ll 1$ differ by several orders of 
magnitude.
Plotting residual saturation against the correct ratio 
of viscous to capillary forces should result in a
breakpoint at $\Ca\approx 1$ \cite{hil94c,dul92}.

We now proceed to show that a partial understanding of
the macroscopic force balance can already be obtained 
from the traditional equations of motion.
The main result of this analysis is a preliminary experimental 
validation of the dimensional considerations introduced 
in \cite{hil94c}.

\section{Discontinuous vs. Continuous Mode Displacement}

Before embarking on a discussion of the force balance
during capillary desaturation it is crucial to emphasize
an important difference in the way capillary desaturation 
curves are measured.
\begin{enumerate}
\item
In the first method of measuring $\Sor(\Ca)$ the oil in
a fully oil saturated sample is displaced with water
at a very low $\Ca$.
After the oil flow at the sample outlet stops
and several pore volumes of water have been injected
without producing more oil the flow rate is increased.
Again the oil flow is monitored until no more
oil appears at the outlet.
In this way the flow rate is increased iteratively 
until $\Sor$ has fallen to zero.
After the first injection the oil configuration is 
discontinuous (or disconnected) and it remains 
disconnected throughout the rest of the experiment.
This mobilization mode will be called discontinuous.
\item
In the second method of measuring capillary desaturation 
curves one
again starts from a fully saturated sample, and
performs a waterflood at a given value of $\Ca$.
After oil flow ceases at the outlet the
residual saturation is determined.
Then the sample is again saturated fully with
oil, a new value of $\Ca$ is chosen, and the
injection is repeated.
In this experiment the injection starts always
with a connected oil phase contrary to the
previous experiment where the oil phase is
discontinuous at higher $\Ca$.
This mobilization mode will be called continuous.
\end{enumerate}
The capillary number required to reach a given
$\Sor$ is known to be much lower in the continuous
mode than in the discontinuous mode
\cite{MCT88,LDS81}.
This is also seen in Figure \ref{ocp} which shows 
continuous mode data and discontinuous mode data for
the unconsolidated glass beads.
From the fact that the equations of motion are only valid in the
subinterval $\Swi<\SS_\wa< 1-\Sor$ it follows that they cannot
be applied to discontinuous mobilization mode experiments.
They should however be valid for the continuous mode to the
extent that the equations themselves are correct.
Therefore we discuss next the macroscopic force balance predicted
by these equations for the continuous mobilization mode.

\section{Macroscopic Force Balance}

It was shown in \cite{hil94c} that the balance of macroscopic
viscous and capillary forces is not represented appropriately
by the microscopic capillary number $\Ca$, and that a new
macroscopic capillary number $\CA$ should be used instead.
Here we generalize those results to anisotropic and inclined
porous media and apply them to replot the $\Sor$ data obtained
for the continuous mode displacement.
To this end we rewrite the equations (\ref{Em5}) and (\ref{Em6})
in dimensionless form.
Introducing the matrix
\begin{equation}
\ML =
\left(
  \begin{array}{ccc}
   L_x & 0 & 0\\
   0 & L_y & 0\\
   0 & 0 & L_z
  \end{array}
\right)
\end{equation}
and defining $\LL$ and $\MLL$ through
\begin{equation}
\LL = ({\rm det} \ML)^{(1/3)}
\end{equation}
\begin{equation}
\ML = \LL\;\MLL
\end{equation}
one defines dimensionless quantities
\begin{equation}
\XX = \ML\;\XXX = \LL\;\MLL\;\XXX
\end{equation}
\begin{equation}
\NAB = \ML^{-1}\NABB = (\LL\;\MLL)^{-1}\NABB .
\end{equation}
Define $\KK$ and $\MKK$ through
\begin{equation}
\MK =
\left(
  \begin{array}{ccc}
   k_{xx} & k_{xy} & k_{xz}\\
   k_{xy} & k_{yy} & k_{yz}\\
   k_{xz} & k_{yz} & k_{zz}
  \end{array}
\right)
\end{equation}
\begin{equation}
\KK = ({\rm det} \MK)^{(1/3)}
\end{equation}
\begin{equation}
\MK = \KK\;\MKK
\end{equation}
The time scale is normalized as
\begin{equation}
\TT = \frac{L_xL_yL_z\ph\;\TTT}{Q}
\end{equation}
using the volumetric flow rate $Q$.
Time is measured in units of injected pore volumes.
The pressure is normalized using the macroscopic 
equilibrium capillary pressure as \cite{hil93h,hil94c}
\begin{equation}
\PP = \PP_b\;\PPP
\label{pressure}
\end{equation}
where 
\begin{equation}
\PP_b = \PP_c\left((S_{\wa i}-S_{\oi r}+1)/2\right)
\label{pnorm}
\end{equation}
is the pressure at an intermediate saturation.
$\PP_c(S_\wa)$ is the equilibrium capillary pressure
function.

With these definitions the dimensionless macroscopic capillary 
numbers for oil and water, defined as 
\begin{equation}
\CA_\wa = \frac{\mu_\wa Q}{\PP_b\KK\LL}
\label{CA}
\end{equation}
\begin{equation}
\CA_\oi = \CA_\wa\frac{\mu_\oi}{\mu_\wa}
\end{equation}
give an expression of the balance between
macroscopic viscous and capillary forces.
The macroscopic gravity numbers
\begin{equation}
\GR_\wa = \frac{\mu_\wa Q}{\rho_\wa g\:\KK\LL^2}
\end{equation}
\begin{equation}
\GR_\oi = \GR_\wa\frac{\mu_\oi}{\mu_\wa}\frac{\rho_\wa}{\rho_\oi}
\end{equation}
express the viscous to gravity force balance.
Finally the gravillary numbers
\begin{equation}
\GL_\wa = \frac{\rho_\wa g\:\LL}{\PP_b}
\end{equation}
\begin{equation}
\GL_\oi = \GL_\wa\frac{\rho_\oi}{\rho_\wa}
\end{equation}
express the ratio between gravitational and capillary
forces.
The well known bond number, measuring the magnitude of buoyancy 
forces, is given as
\begin{equation}
\Bo = \GL_\wa - \GL_\oi
\end{equation}
in terms of the gravillary numbers.

With these definitions
the dimensionless equations of motion may be rewritten as
\begin{equation}
\frac{\ds \partial \SS_\wa}{\ds \partial \TTT} =
\NABB \cdot \left\{\MAA\;\kwr
\left[\CA^{-1}_\wa\NABB \PPP_\wa -\GR^{-1}_\wa\;\ggg\right]
\right\}
\label{Em9}
\end{equation}
\begin{equation}
-\frac{\ds \partial \SS_\wa}{\ds \partial \TTT} =
\NABB \cdot \left\{\MAA\;\kor
\left[\CA^{-1}_\wa\NABB (\PPP_\wa+\PPP_c) -\GR^{-1}_\oi\;\ggg\right]
\right\}
\label{Em10}
\end{equation}
where the dimensionless matrix
\begin{equation}
\MAA = \MLL^{-1}\MKK \MLL^{-1}=
\left(
  \begin{array}{ccc}
   \frac{\ds \LL^2 k_{xx}}{\ds L_x^2 k} & 
   \frac{\ds \LL^2 k_{xy}}{\ds L_xL_y k} & 
   \frac{\ds \LL^2 k_{xz}}{\ds L_xL_z k}\\[12pt]
   \frac{\ds \LL^2 k_{xy}}{\ds L_xL_y k} & 
   \frac{\ds \LL^2 k_{yy}}{\ds L_y^2k} & 
   \frac{\ds \LL^2 k_{yz}}{\ds L_yL_z k}\\[12pt]
   \frac{\ds \LL^2 k_{xz}}{\ds L_xL_z k} & 
   \frac{\ds \LL^2 k_{yz}}{\ds L_yL_z k} & 
   \frac{\ds \LL^2 k_{zz}}{\ds L_z^2 k}
  \end{array}
\right)
\label{aspectr}
\end{equation}
contains generalized ``aspect ratios''.
The vector
\begin{equation}
\ggg^T = (\MLL\;\MR\;\EE)^T = 
\left(-\frac{L_x}{\LL}\sin\dip_y,0,-\frac{L_z}{\LL}\cos\dip_y\right)
\end{equation}
represents the effect of dip angle and geometric shape of the system
on the gravitational driving force.

\section{Application to Experiment}
We are now in a position to plot the capillary desaturation curves
against the macroscopic capillary number $\CA$ which represents
the balance between macroscopic viscous and capillary forces.
To do so we have searched the literature for capillary desaturation
measurements and found Refs.
\cite{wil86,lak89,DB54,tab69,TKS73,DTL69,EHR74,lef73,GT79,LDS81}.
Unfortunately none of the publications contains all the necessary
flow and medium parameters to calculate $\CA$. 
Measuring all the flow parameters for a displacement process
is costly and time consuming, and hence they are rarely
available (see also \cite{hil96a}).
While permeability, porosity and the fluid parameters
such as vicosities and surface tensions are usually
available capillary pressure data, relative permeabilities
and residual saturations are not routinely measured.
Hence we extract the required parameters from different 
publications assuming that they have been measured 
correctly and are reproducible anywhere and at all 
times.
In spite of all the uncertainties it is well known that 
the capillary pressure curves of unconsolidated sands 
with various grain sizes can be collapsed using the 
Leverett-$j$-correlation \cite{LLT42,bea72,DS74}.
The Leverett-$j$-correlation states essentially that
\begin{equation}
\PP_c(\SS_\wa) = \sigma_{\oi\wa}\sqrt{\frac{\ph}{k}}\;j(\SS_\wa) .
\label{leverettj}
\end{equation}
Often the formula contains in addition an average contact angle 
at a three phase contact.
We do not include the wetting angle as it is generally
unknown, and including it would not change our results significantly.
Similar to the capillary pressure data, the $\Sor$ data for 
unconsolidated sands seem to be well established \cite{WM85,MCT88}
in spite of larger fluctuations of the results.
Because most $\PP_c$- and $\Sor$-data are available for 
unconsolidated sand and standard sandstones such as Berea 
or Fontainebleau we limit our analysis to these two cases.

The experimental $\Sor$ data analyzed here are taken from
Ref. \cite{abr75} for sandstones and from Ref. \cite{MCT88}
for unconsolidated glass beads.
In Figure \ref{ocp} we show the capillary desaturation curve
for oil-water displacement in a typical sandstone 
(sample No. 799 from \cite{abr75}) using star symbols.
These data were obtained in the continuous mode of displacement.
The values of the surface tension and fluid viscosities
for this experiment are given in Table \ref{table}.
We also show continuous mode $\Sor$-data for unconsolidated
glass beads from Figure 6 in \cite{MCT88}.
To calculate $\CA$ we have used the values
given in Table \ref{table}.

Plotting the data against $\CA$ we obtain Figure \ref{ncp}.
It is seen that the continuous mode displacements give
a breakpoint for $\CA\approx 1$ while the discontinuous
mode displacements have their breakpoint at a higher value.
This is consistent with the idea that the traditional
equations of motion (\ref{Em5}) and (\ref{Em6}) should
be applicable to the continuous mode but not to the
discontinuous case.

The result obtained here is consistent with the theoretical 
predictions from \cite{hil94c}.
To fully validate our use of $\CA$ as a correlating group
for plotting continuous mode $\Sor$ data, however, it would
be desirable to vary $\CA$ by varying the system size $\LL$.
If the $\Sor$ curves obtained for different media and 
length scales $\LL$ also show their breakpoint at $\CA\approx 1$
this would give further evidence for the applicability of
the traditional equations of motion.
We consider it possible, however, that deviations appear
indicating a breakdown of the equations also for continuous
mode displacement \cite{hil98b,BH98}.

As stated above the equations of motion are not applicable
to discontinuous mode displacements because of the constraint
$\Swi<\SS_\wa<1-\Sor$.
Nevertheless we can use the group $\CA$ to estimate an upper\
bound for the size of residual blobs.
As the flow rate is increased the residual blobs whose size is
so large that the viscous drag forces on them exeed the
capillary retention forces will break up and coalesce with 
other blobs which may again break up and coalesce further
downstream \cite{WM85,CKM88}.
The condition that the viscous forces dominate
the capillary forces $\CA\geq 1$ predicts that after
a flood with $\CA$ the porous medium contains only blobs 
of linear size smaller than 
\begin{equation}
\LL_{blob} \leq \frac{\mu_\wa Q}{\PP_b\KK} .
\label{blob}
\end{equation}
This result is of importance for microscopic
models \cite{AH98} of breakup and coalescence 
during immiscible displacement.

ACKNOWLEDGEMENT:
One of us (R.H.) thanks Dr. P.E. {\O}ren for many useful
discussions.
We are grateful to the Deutsche Forschungsgemeinschaft
for financial support.
\newpage
\begin{table}
\setdec 0.00
\caption{The sample and flow parameters used in the calculations.
The permeabilities for the glass bead packs were obtained using 
the relation between their radius and permeabilities published 
in \protect\cite{lak89} on page 47.
The capillary pressure was estimated using the Leverett-$j$-function
correlation (\protect\ref{leverettj}).
For sandstone the $j$-function from
Ref. \protect\cite{DS74} was used and for bead packs
the one from Ref. \protect\cite{LLT42} was used. 
The $j$-function was evaluated close to 
$(S_{\wa i}-S_{\oi r}+1)/2$ according to 
(\protect\ref{pnorm}).}
\vspace*{3cm}
\begin{tabular}{l|c|c|c|c}
     & permeability & porosity  & surface tension & cap. pressure\\
sample & $k$ ($10^{-12}$m$^2$) & $\ph$ & $\sigma_{\oi\wa}$ (N/m) & $\PP_b$ (Pa)\\
\tableline\tableline
sandstone 799 (Ref.\protect\cite{abr75})
&   $0.14$ (Ref.\protect\cite{abr75}) 
& $0.28$ (Ref.\protect\cite{abr75}) 
& $3.37\times10^{-2}$ (Ref.\protect\cite{abr75}) 
& $1.7\times10^{4}$ (Ref.\protect\cite{DS74})\\
\tableline
$70\mu m$ bead pack (\protect\cite{MCT88})
& $10$ (Ref.\protect\cite{lak89})
& $0.36$ (Ref.\protect\cite{lak89,WKT84})
& $2.8\times10^{-3}$ (Ref.\protect\cite{MCT88})
& $2.4\times10^{2}$ (Ref.\protect\cite{LLT42})\\
$115\mu m$ bead pack (\protect\cite{MCT88})
& $30$ (Ref.\protect\cite{lak89})
& $0.36$ (Ref.\protect\cite{lak89,WKT84}) 
& $1.18\times10^{-2}$ (Ref.\protect\cite{MCT88})
& $5.8\times10^2$ (Ref.\protect\cite{LLT42})\\
\end{tabular}
\label{table}
\end{table}
~
\newpage
\section*{Figure Captions}

\newcounter{fig}
\begin{list}{\textbf{Figure \arabic{fig}:}}
{\usecounter{fig}
\setlength{\labelwidth}{2.2cm}
\setlength{\labelsep}{0.3cm}
\setlength{\itemindent}{0pt}
\setlength{\leftmargin}{2.5cm}
\setlength{\rightmargin}{0cm}
\setlength{\parsep}{0.5ex plus0.2ex minus0.1ex}
\setlength{\itemsep}{0ex plus0.2ex}}
\item \label{ocp}
Experimentally measured capillary desaturation curves
(capillary number correlations)
for bead packs \cite{MCT88} (solid lines with circles and squares)
and sandstone \cite{abr75} (star symbols)
as a function of microscopic 
capillary number $\Ca=\mu_\wa v/\sigma_{\oi\wa}$.
The dashed line is the desaturation curve for continuous
mode displacement.
The dash-dotted line marks the plateau value for
sandstone.
\item \label{ncp}
Same as Figure \ref{ocp} but plotted against the
macroscopic capillary number
$\CA=\mu_\wa Q/(\PP_b\KK\LL)$ from eq. (\ref{CA}).
Note that the breakpoint for continuous mode displacement
occurs around $\CA\approx 1$.
\end{list}
\newpage
\voffset-2cm 
{\small\sffamily 
A. Lucian and R. Hilfer \hfill Figure \ref{ocp}}\\[3cm]
\psfig{figure=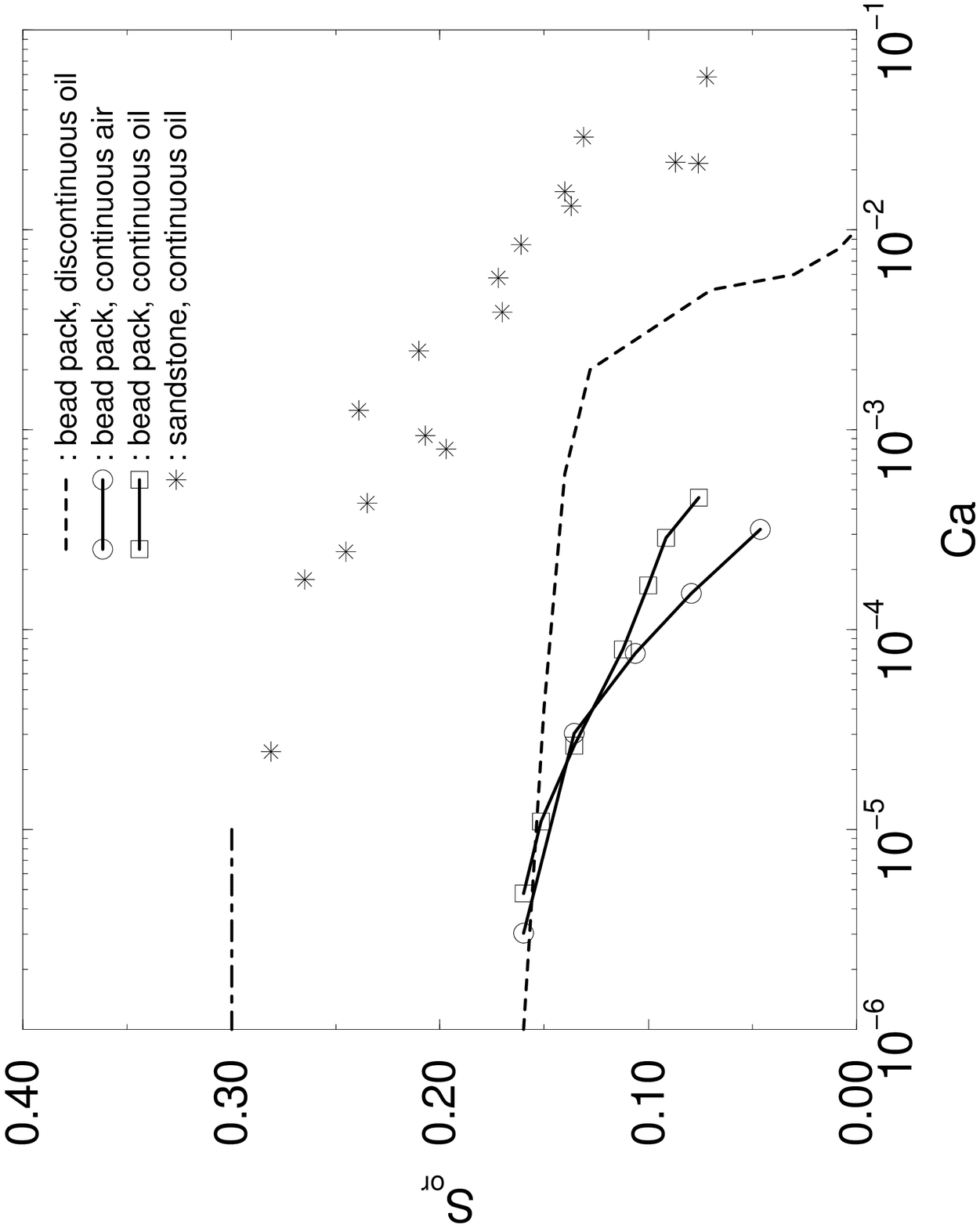,angle=0,width=16cm}

\newpage
\voffset-2cm 
{\small\sffamily 
A. Lucian and R. Hilfer \hfill Figure \ref{ncp}}\\[3cm]
\psfig{figure=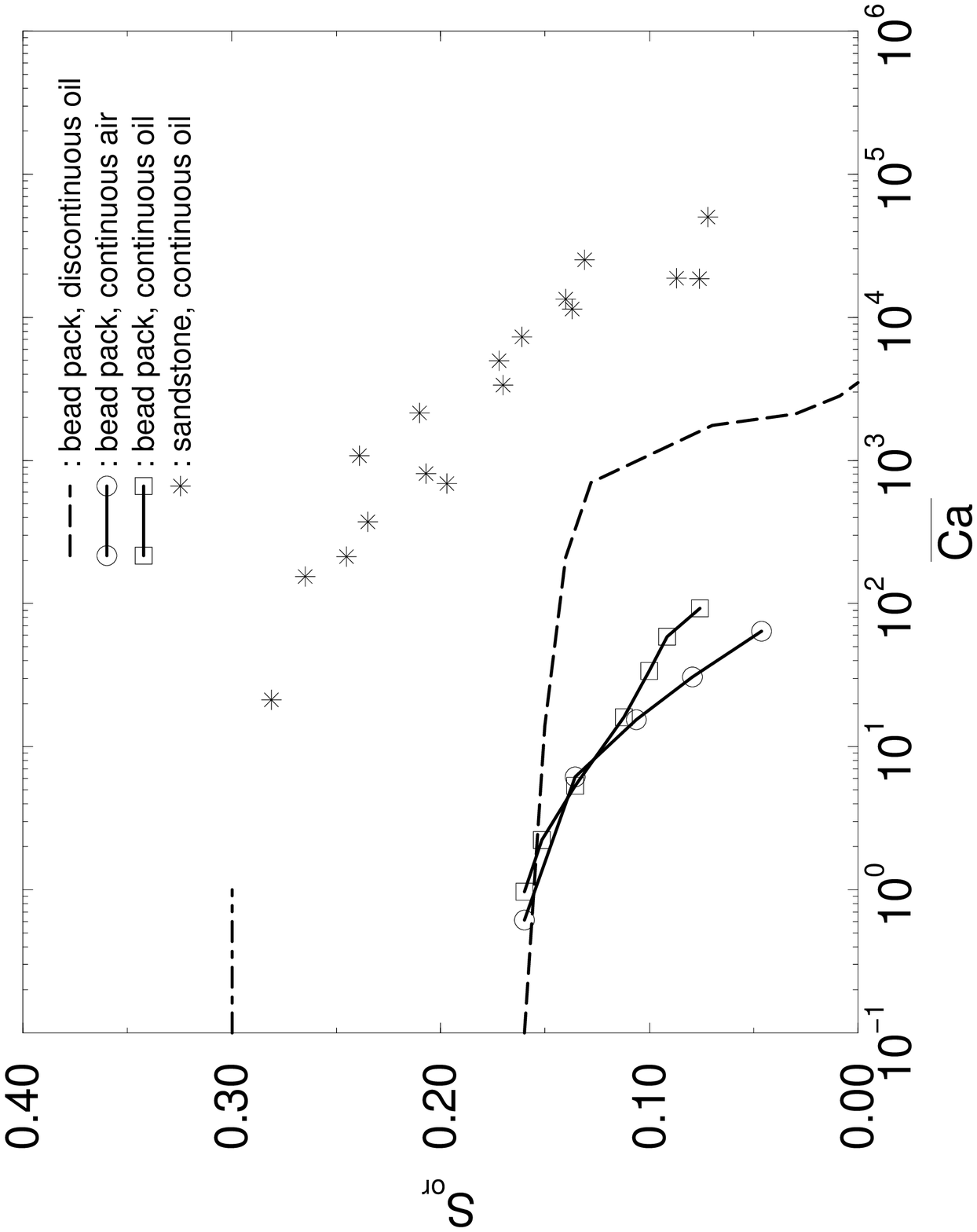,angle=0,width=16cm}

\newpage
\voffset0cm
\newpage


\end{document}